# Optimized Interpolations and Nonlinearity

# In Numerical Studies of Woodwind Instruments

Apostolos Skouroupathis, Haralambos G. Panagopoulos

University of Cyprus, Department of Physics, Kallipoleos 75, P.O.Box 20537, CY-1678, Nicosia, Cyprus, {**php4as01,haris**}@ucy.ac.cy

We study the impedance spectra of woodwind instruments with arbitrary axisymmetric geometry. We perform piecewise interpolations of the instruments' profile, using interpolating functions amenable to analytic solutions of the Webster equation. Our algorithm optimizes on the choice of such functions, while ensuring compatibility of wave fronts at the joining points. Employing a standard mathematical model of a single-reed mouthpiece as well as the time-domain reflection function, which we derive from our impedance results, we solve the Schumacher equation for the pressure evolution in time. We make analytic checks that, despite the nonlinearity in the reed model and in the evolution equation, solutions are unique and singularity-free.

## 1     Introduction

In this work we study the impedance spectrum and the time evolution of radiated pressure, for wind instruments with a single reed. The subject has been studied widely in the literature, from a number of different points of view (for a recent overview of different approaches, see [1]). Regarding the impedance spectrum, our focus is on different interpolations of the horn's profile. More specifically we are interested in interpolations that lead to analytic solutions of the Webster equation, without any mismatch of the wave fronts at segment boundaries. We consider different criteria for establishing an optimal interpolation and compare our results with standard piecewise conical approximations. In particular, we examine the effects of the interpolation on impedance and on resonance frequencies, presenting also the temperature dependence of our results. We use a Dolnet clarinet as our prototype example.

To test our results on impedance, we apply them to the calculation of radiated pressure as a function of time. This relies on solving numerically the Schumacher equation [2], which is a convolution-type equation in real time. Input ingredients for this equation are the reflection function in the time domain, which can be derived from the numerical results on the impedance spectrum, as well as a mathematical description for the mouthpiece. Such a description is provided by a widely used model [3], applicable to single reed instruments.

Given that the model is nonlinear, we carry out an analytic investigation of the time evolution, to ensure that we are not led to any unphysical singularities or bifurcations.

The outline of the paper is as follows: Section 2 sets up basic definitions and notation, and gives the various analytical solutions to the Webster equation, together with the corresponding impedance ratios. Section 3 discusses an optimal interpolation and presents results of the numerical calculations, showing the effect of temperature and of different interpolations on the impedance spectrum. Section 4 applies the impedance results to the calculation of radiated pressure. It presents numerical results for the reflection function in the time domain, comparing different interpolations, and results on pressure, as a function of time.

(A longer write-up, including a more complete list of references, can be found at: http://xxx.lanl.gov/)

## 2     The Webster equation-Analytic solutions, impedance ratios

We start with some standard definitions and notation [1,4]; the results presented in this Section, are for the most part, well known in the literature. The wave equation for air pressure $p(\vec{r},t)$ is:

$$\frac{\partial^2 p}{\partial t^2} = c^2 \nabla^2 p \qquad (1)$$

For typical values of the temperature T ($^o$C) and "standard" air composition, the speed of sound $c$ and the density of air $\rho$ behave as:

$$c = 332\ (1 + 0.00166\ T)\ m/s \qquad (2)$$
$$\rho c = 428\ (1 - 0.0017\ T)\ kg/m^2/s \qquad (3)$$

In a Fourier description, the pressure is a superposition of components with time dependence $\exp(j\omega t)$, where the spatial part obeys:

$$\nabla^2 p + k^2 p = 0 \qquad (4)$$

with $k=\omega/c$. The flow velocity $\vec{u}(x,t)$ follows:

$$\vec{u}(x,t) = \frac{j}{\rho\omega}\vec{\nabla}p \quad (5)$$

Integrating $\vec{u}(x,t)$ over a surface, typically a wave front, gives the volume flow $U(t)$. The effect of thermal conduction and viscosity is commonly incorporated into a redefinition of the wave number $k$, in terms of a phase velocity $u$ and an attenuation factor $\alpha$: $k \to \omega/u - j\alpha$

In a musical instrument, viscosity and heat conduction are localized on boundary layers of thickness $\delta_v$ and $\delta_t$, respectively. Their approximate effect on $u$ and $\alpha$ is:

$$u \approx c\left[1 - \frac{\delta_v + (\gamma-1)\delta_t}{a\sqrt{2}}\right], \quad \alpha \approx \frac{\omega}{c}\left[\frac{\delta_v + (\gamma-1)\delta_t}{a\sqrt{2}}\right] \quad (6)$$

where $\gamma = C_p/C_v$ and $a$ is the pipe radius. Near 300°K :

$$\delta_v \approx \left[632.8 f^{1/2}(1 - 0.0029\Delta T)\right]^{-1} \quad (7)$$

$$\delta_t \approx \left[532.8 f^{1/2}(1 - 0.0031\Delta T)\right]^{-1} \quad (8)$$

where $f = \omega/2\pi$ is the frequency and $\Delta T$ is the temperature deviation from 300°K.

Inside a horn of arbitrary axisymmetric shape, the variation of pressure is frequently taken to be longitudinal, with wave fronts which are axisymmetric surfaces perpendicular to the walls; these surfaces are essentially spherical (exactly so, for a conical horn). Denoting by $S(x)$ the surface area of a wave front intersecting the axes at point $x$, the wave equation reduces to the so-called Webster equation:

$$\frac{1}{S}\frac{\partial}{\partial x}\left(S\frac{\partial p}{\partial x}\right) = \frac{1}{c^2}\frac{\partial^2 p}{\partial t^2} \quad (9)$$

This equation cannot be analytically solved for arbitrary profiles $f(x)$; typically one must separate the horn into a sequence of segments, and interpolate each segment with a profile which is amenable to an analytic solution.

There are three types of horn profiles solving the Webster equation analytically:

i) Conical profile: $a(x) = a_0 + Tx$

This is the most widely studied case. The two independent solutions of the Webster equation are:

$$p_1(x) = e^{-jkx}/(a_0 + Tx), \quad p_2(x) = e^{+jkx}/(a_0 + Tx) \quad (10)$$

Since this profile depends on two parameters only, using it as an interpolation in a segment will match the actual profile, but not its derivative, at the endpoints of the segment. This will lead to an undesirable mismatch of wave fronts in adjacent segments.

ii) "Bessel" profile: $a(x) = b(x - x_0)^{-\varepsilon} \equiv b\Delta x^{-\varepsilon}$

The two independent solutions are now Bessel functions of the first and second kind:

$$p_1(x) = \Delta x^{\varepsilon + 1/2} J_{\varepsilon + 1/2}(k\Delta x), \quad p_2(x) = \Delta x^{\varepsilon + 1/2} Y_{\varepsilon + 1/2}(k\Delta x) \quad (11)$$

The presence of a third parameter in the profile allows for derivative matching and thus closes the gap between adjacent wave fronts. Nevertheless, this functional form is too restrictive to fit most realistic profiles, requiring, e.g., that both the profile and its slope be monotonic.

iii) "Transcendental" profile:

By this we denote profiles having either a sinusoidal or hyperbolic form:

$$a(x) = a_0(\cos mx + T \sin mx) \quad (12a)$$

$$a(x) = a_0(\cosh mx + T \sinh mx) \quad (12b)$$

The solutions now read:

$$p_1(x) = e^{-j\sqrt{k^2 \pm m^2}\,x}/a(x), \quad p_2(x) = e^{+j\sqrt{k^2 \pm m^2}\,x}/a(x) \quad (13)$$

The positive (negative) sign in front of $m^2$ corresponds to the sinusoidal (hyperbolic) case. While a sinusoidal form cannot be applied to a horn profile as a whole, it is indispensable for the description of concave segments ($f''(x) < 0$), which are typically present.

The volume flow $U(x)$ and impedance $Z$ are given by:

$$U(x) = S(x)u(x) = \frac{j}{\rho k c}S(x)\frac{dp}{dx}, \quad Z(x) = p(x)/U(x) \quad (14)$$

Writing the pressure in terms of the two solutions:

$$p(x) = \gamma_1 p_1(x) + \gamma_2 p_2(x) \quad (15)$$

the ratio of impedances at the left and right ends of a segment, $x_L, x_R$, is:

$$Z(x_L)/Z(x_R) = \frac{\left[p(x)/p'(x)S\right]_{x=x_L}}{\left[p(x)/p'(x)S\right]_{x=x_R}} \quad (16)$$

Only the ratio $\gamma_2/\gamma_1$ is unknown in the above relation. It is fixed by impedance matching: If $Z$ at the $(i+1)^{th}$ segment is known, then $\gamma_2/\gamma_1$ at the $i^{th}$ segment is:

$$-\frac{j\rho k c}{S}\frac{p_1^i + (\gamma_2^i/\gamma_1^i)p_2^i}{\left[p_1^i + (\gamma_2^i/\gamma_1^i)p_2^i\right]'}\bigg|_{x=x_0} = Z^{i+1}(x_0) \quad (17)$$

At the free edge of the rightmost segment, one possible standard approximation to the impedance corresponds to the presence of a plane flange, leading to:

$$Z^{fl} = Z_0\left(1 - J_1(2ka)/ka\right) + jZ_0 H_1(2ka)/ka \quad (18)$$

where $J_1$ and $H_1$ are Bessel and Struve functions, and $Z_0$ is the characteristic impedance, $Z_0 = \rho c/S$. Other alternatives to $Z^{fl}$ are possible, but a correct expression for a horn with a thin flange is not available.

Thus, to calculate the input impedance $Z_{IN}$ of a horn we iterate over segments, starting from the flare (right end), once a suitable interpolation for each segment has been established. Impedance matching on the right edge of a segment gives the ratio $\gamma_2/\gamma_1$ via Eq.(17), while the impedance at the left edge is obtained through Eq.(16).

Deciding on the optimal interpolation for each segment requires a detailed investigation, which we address next.

# 3  Optimized interpolations - Results on impedance

The profile of a horn is typically described though a set of values of the horn radius corresponding to selected locations along the horn's symmetry axis. At all other locations, the profile must be smoothly interpolated. The general problem of interpolation is certainly a widely studied area in Applied Mathematics, where standard sets of functions (polynomial splines, Padé approximants, etc.) are typically used; the difference in this context is the requirement to use only those functional forms which lead to analytic solution of the Webster equation, as described in Section 2.

The simplest approximation is the conical one, in which the profile is made up of adjacent straight line intervals, joined together at their endpoints; the horn is thus a concatenation of straight conical sections. Despite the mismatch in wave fronts, this approximation is the most standard one and will be used as a point of comparison in testing other interpolations.

The remaining available functional forms (Bessel, transcendental) contain an extra (third) free parameter; two parameters are then used up in order to match the prescribed values of the radius at the two endpoints of a segment and the third parameter can be tuned so as to affect the slope of the profile. There are many ways of adjusting the slope and one must use some criterion in order to select the optimal one. One family of algorithms which we tested involves visiting each segment in sequence, and matching its slope to that of the previous segment, at their common endpoint. Thus, if $f_L$ and $f_R$ are the prescribed values of the profile at the two endpoints $x_L$, $x_R$, of a segment, and $f'_R$ is a prescribed value of the slope at the right end, we have three equations to solve for the three parameters.

## 3.1  Applicability of the Bessel and transcendental interpolations

Let us examine these equations for the Bessel interpolation, in order to test the limits of its applicability:

$$b(x_L - x_0)^{-\varepsilon} = f_L, \quad b(x_R - x_0)^{-\varepsilon} = f_R, \quad -\varepsilon b(x_R - x_0)^{-\varepsilon-1} = f'_R \quad (19)$$

We must check for which values of $f_L$, $f_R$, $f'_R$, these equations admit a (unique) solution for the parameters $b, \varepsilon, x_0$. We find this to be the case provided:

$$\frac{\ln(f_R/f_L)}{(x_R - x_L)(f'_R/f_R)} > 1 \quad (20)$$

If this criterion is not fulfilled for each segment, the Bessel interpolation is not applicable.

A similar procedure can be applied to sinusoidal interpolations. The equations to be solved now read:

$$a_0 (\cos m x_{L,R} + T \sin m x_{L,R}) = f_{L,R} \quad (21)$$

$$a_0 m (-\sin m x_R + T \cos m x_R) = f'_R \quad (22)$$

Again, we must check under which conditions there will be a solution for $a_0$, $m$, $T$. The above can be combined to give an equation for $m$  ($\Delta x = x_R - x_L$) :

$$f'_R \frac{\sin m \Delta x}{m \Delta x} = \frac{f_R \cos m \Delta x - f_L}{\Delta x} \quad (23)$$

(Similarly for hyperbolic interpolations, replacing sin, cos by sinh, cosh) We can now distinguish two cases:
a) If $f'_R < (f_R - f_L)/\Delta x$, then the l.h.s. of Eq.(23) is smaller than the r.h.s. for $m=0$ and larger for $m=\pi/\Delta x$. Clearly, then, there will be one solution of the form $a_0(\cos mx + T\sin mx)$, in the acceptable range for m.
 b) If $f'_R > (f_R - f_L)/\Delta x$, then it is the hyperbolic analogue of Eq.(23) which will have one solution.

## 3.2  Improved interpolations

The algorithms described above, although they lead to a continuous profile with continuous first derivative throughout the horn's length, can be unstable, since a small change in the slope of one segment may propagate and get amplified in the slopes of subsequent segments, leading to the appearance of unphysical deformations in the horn's profile. We have tried a number of variants of this approach, starting from either the mouth or the throat of the horn, or from a middle segment.

Next we tried a variant of the above algorithms which is parallel, rather than sequential, in the segments, and thus avoids any accumulation of deformations. The basic idea in this variant is to render each segment more and more concave or convex in a gradual manner, so as to realize a better match at end points. The procedure stops when further improvement of matching in one endpoint spoils matching at the other endpoint. This variant, while it avoids accumulation of any unphysical deformations, does not lead to an exact matching of wave fronts, especially near inflection points.

Our final interpolation algorithm, at the expense of being more complicated, avoids the above problems: It is parallel and leads to exact derivative matching throughout. In particular, we take a weighted average of the straight line slopes on both sides of a junction point and require that the derivatives of the interpolating functions on either side of the junction match this average. Our algorithm, in a nutshell, works as follows:

▪ Each segment [$x_L$, $x_R$] is split into two subsegments at some intermediate point $x_0$, to be determined. In each subsegment, a hyperbolic (or sinusoidal, according to a certain criterion) interpolating function $F_{L,R}(x)$ is used.

▪ Matching of profile values and derivatives is required at $x_L$, $x_0$, $x_R$. This amounts to six conditions on the six interpolating parameters.

- The correct choice of hyperbolic versus sinusoidal interpolating function can be elucidated based on Figure 1: We denote by $f'$ the straight line slope in the segment $[x_L, x_R]$, and by $f'_L$, $f'_R$ the weighted slopes at the left and right endpoints, which must be matched by our interpolating functions. The following cases arise:

a) $f'_L > f'$, $f'_R > f'$ : $F_L(x)$ sinusoidal, $F_R(x)$ hyperbolic

b) $f'_L < f'$, $f'_R < f'$ : $F_R(x)$ sinusoidal, $F_L(x)$ hyperbolic

c) $f'_R > f' > f'_L$ : This is the case shown in Figure 1. The segments with slopes $f'_L, f'_R$ are extended until they intersect at some point $x_I$. Let us consider the case $x_I > x_0$. $F_L(x)$ and $F_R(x)$ will be joined at some point which will certainly be between $P$ and $P'$, if their derivatives are to be equal, since $F'_L(x) - F'_R(x) < 0 \ (>0)$ if $F_L(x)$ and $F_R(x)$ are made to pass through $P$ ($P'$). In all cases, $F_R(x)$ lies above the segment with slope $f'_R$, and is thus hyperbolic. To determine what type of function $F_L(x)$ is, we must check whether $F_L(x)$ and $F_R(x)$ will meet above or below $P''$: If $F_R(x)$ (when it is made to pass through $P''$) satisfies $F'_R(x_0) > f'_L$, then $F_L(x)$ and $F_R(x)$ will have to be joined above $P''$ and consequently $F_L(x)$ will be hyperbolic, otherwise it will be sinusoidal. Similar considerations apply to the symmetric cases: $x_I < x_0$, $f'_R < f' < f'_L$.

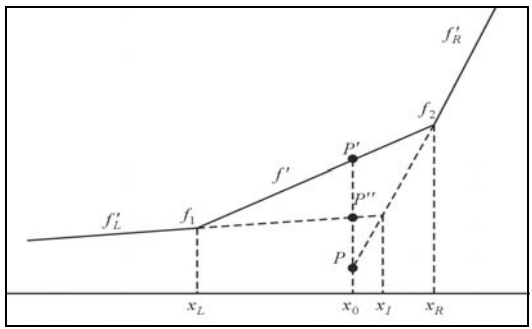

Figure 1: Criteria for the choice of interpolating functions

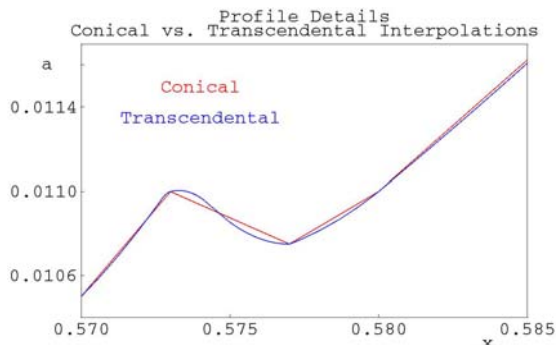

Figure 2: Comparison of the profile's details for a Dolnet type clarinet, around a junction point, using the conical and transcendental interpolations (radius a vs. axis coordinate x, both measured in meters)

Our algorithms for determining the optimal inter-polations have been worked out in Mathematica, and are available from the authors upon request. They have the form of a module which takes as input a list of profile values at different points along the horn's axis, and returns lists of interpolating functions for each segment. The difference between the conical and our optimized transcendental interpolation is exemplified in Figure 2, which shows a detail from the profile of a Dolnet type clarinet, around a junction point.

### 3.3 Results on impedance

All of our results refer to a Dolnet type of clarinet. Figure 3 exhibits the spectrum of the impedance using the conical approximation. Figure 4 gives a comparison of conical and exponential approximations, as regards impedance. Differences in impedance are relatively small, of the order of 2% (10%) for the real (imaginary) part, in the audible range; resonance frequencies in the two approximations are practically identical (< 0.5 Hz).

Figure 5 displays results on impedance, for conical and transcendental approximations. The frequency range shown is well beyond the audible threshold, but the functional behavior of impedance at these frequencies is important for the reflection function. Figure 6 illustrates the impedance as a function of frequency and temperature. The effect of these parameters is viewed clearly from a top view, as in Figure 7. One observes that impedance maxima are displaced with temperature, and more so at higher frequencies. Figure 8 exhibits the variation of lowest resonances with temperature.

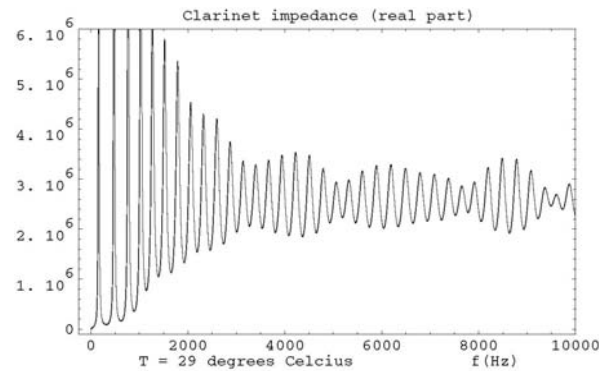

Figure 3: Impedance spectrum (real part, SI units) for a Dolnet clarinet, in the conical approximation

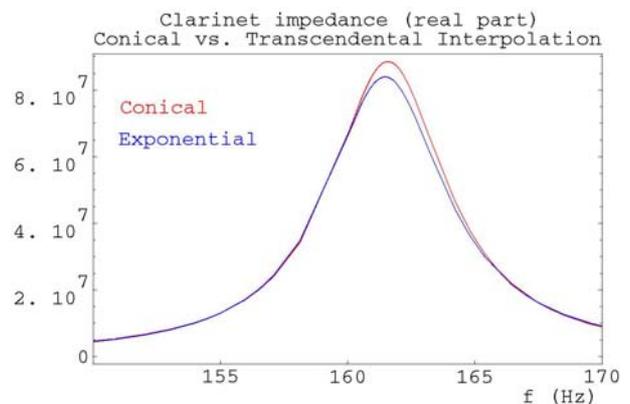

Figure 4: Impedance spectrum (real part, SI units), for the conical and exponential approximation

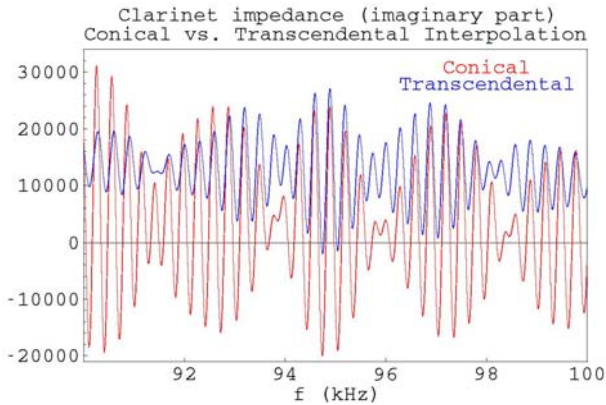

Figure 5: Impedance (imaginary part, SI units) as a function of frequency, for conical and transcendental approximations, in the range 90-100 kHz. This range is well beyond the audible threshold, but the functional behavior of impedance at these frequencies is important for the reflection function in the time domain

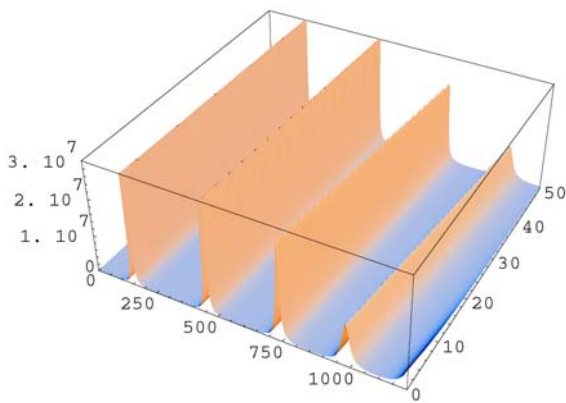

Figure 6: Impedance (real part, SI units) of a Dolnet clarinet as a function of frequency and temperature

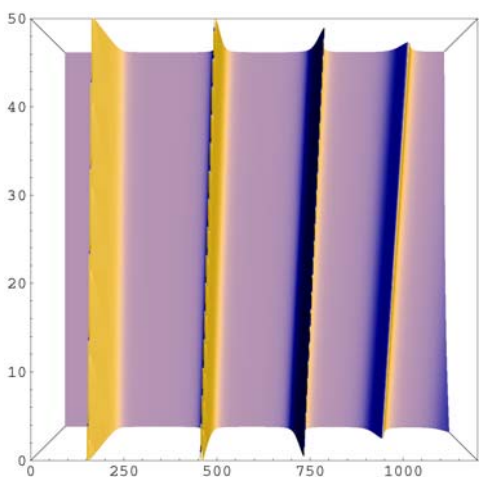

Figure 7: Clarinet impedance (real part, SI units) as function of frequency and temperature. From this view one can observe the displacement of impedance maxima for different temperatures

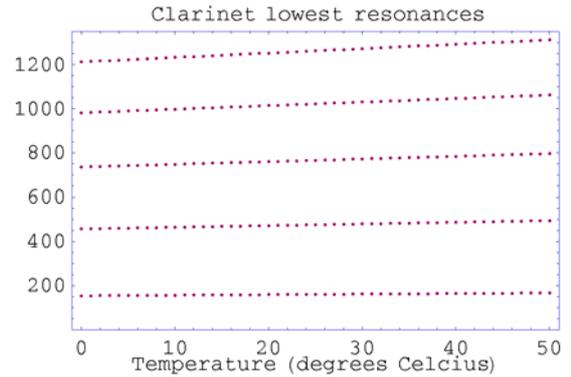

Figure 8: Variation of lowest resonances (in Hz), as a function of temperature

## 4    Mouthpiece model – Reflection function – Schumacher's equation

We employed a standard mathematical description of a single-reed mouthpiece [3], which is quite realistic and simple. Some other recent investigations are: [5,6] and references therein. The model is seen in Figure 9.

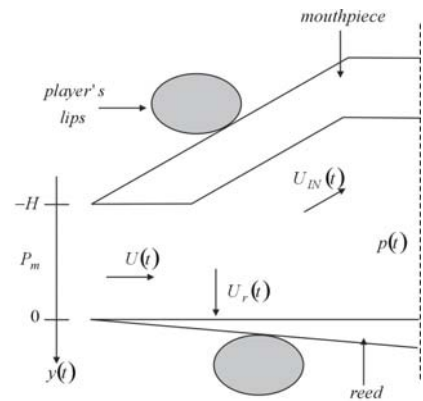

Figure 9: Mouthpiece of a single-reed woodwind

If $P_m$ is the static pressure in the player's mouth, the acoustic flow $U(t)$ entering the instrument is related to the pressure difference $P_m - p(t)$ by Bernoulli's law

$$|P_m - p(t)| = \frac{1}{2}\rho\left\{\frac{U(t)}{\varpi[y(t)+H]}\right\}^2 \quad (24)$$

($\rho$: air density, $\varpi$: width of the reed, $H$: height of the air gap at rest, $y(t)$: position of the reed's extremity). When the reed is completely closed: $U(t)=0$, while $y(t) \geq -H$ must hold at all times. The acoustic flow is then:

$$U(t) = \sqrt{\frac{2}{\rho}}\varpi\sqrt{|P_m - p(t)|}[y(t)+H]\,\text{sign}(P_m - p(t)) \quad (25)$$

The total acoustic flow $U_{IN}(t)$ entering the bore depends on $U(t)$ and on the acoustic flow $U_r(t)$ produced by the reed movement ($S_r$ : effective area of the reed)

$$U_r(t) = S_r \frac{dy(t)}{dt}, \quad U_{IN}(t) = U(t) - U_r(t) \quad (26)$$

Under standard simplifying assumptions, the equation of motion for the reed is ($g_r$, $\omega_r$, $\mu_r$: fixed parameters):

$$\frac{d^2 y(t)}{dt^2} + g_r \frac{dy(t)}{dt} + \omega_r^2 y(t) = \frac{p(t) - P_m}{\mu_r} \quad (27)$$

The pressure is a convolution between $U_{IN}(t)$ and $r(t)$:

$$p(t) = Z_0 U_{IN}(t) + \int_0^\infty r(t')[Z_0 U_{IN}(t-t') + p(t-t')]dt' \quad (28)$$

$r(t)$ is the Fourier transform of the reflection function: $r(\omega) = (Z(\omega) - Z_0)/(Z(\omega) + Z_0)$ and $Z_0 = \rho c/S$. The short range of $r(t)$ implies that only recent values of $p(t)$ contribute to the value of $p(t)$ in the present.

Our algorithm for time evolution proceeds as follows:

▪ At time $t = t_1$ (and at all previous times $0 \leq t < t_1$) the values of $U_{IN}(t_1)$, $U_r(t_1)$, $U(t_1)$, $p(t_1)$ and $y(t_1)$ are known

▪ From Eqs.(26), (27) we find $U_r(t_2)$, $y(t_2)$ ($t_2 = t_1+\Delta t$)

▪ Substituting in Eq.(28) we calculate $p(t_2)$

For the correct functioning of the algorithm we must investigate which are the proper initial conditions, compatible with the physical model. We find for $p(0)$:

$$p(0) = Z_0 \sqrt{\frac{2}{\rho}}\varpi\sqrt{|P_m - p(0)|}H \quad (29)$$

The nonlinearity in the reed model raises the possibility of singularities and bifurcations in the solutions. We have performed a thorough analytic check that such pathologies will never be present. Figure 10 compares the conical and transcendental interpolations, regarding our results for $r(t)$. A smoother behavior is observed coming from the transcendental interpolation. Figure 11 displays the pressure evolution in time; one can observe transient phenomena at initial times. Differences among various interpolations are not perceptible on this scale.

While the observed differences among the various types of interpolation which we have examined are small, they are expected to be more pronounced in pipes of more complicated geometry. In particular, it is important to examine transverse variations in pressure and to carry over a similar approach to pipes with bends. Similarly, the effect of finger holes should be dealt with in a more precise way than standard treatments. A full fledged three dimensional treatment using piecewise interpolating functions, which are compatible with solutions of the 3-d wave equation, would be worth a detailed investigation.

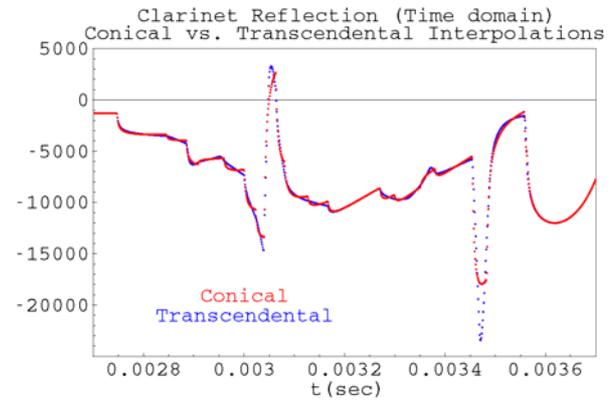

Figure 10: $r(t)$, for a Dolnet clarinet. Comparison between conical and transcendental interpolation

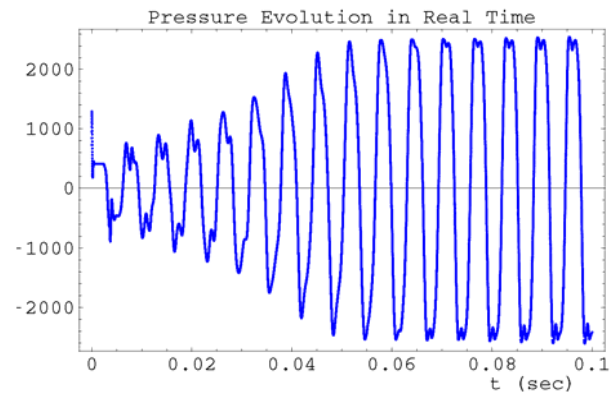

Figure 11: Pressure evolution in time (SI units). The graph shows the transients at initial times